\input{aipcheck.tex}

\def\selectedoptions{final}

\documentclass[
   \selectedoptions
  ]
  {aipproc}

\layoutstyle{6x9}

\SetInternalRegister\hbadness{8000} 

\newcommand\doingARLO[2][]{%
  \ifx\mmref\undefined #1\else #2\fi
}

\begin{document}

\title 
      {Hadron 2001 Conference Summary: Theory}


\author{T.Barnes}{
address={Physics Division, Oak Ridge National Laboratory, 
Oak Ridge, TN 37831-6373, USA \\ 
Department of Physics and Astronomy, University of Tennessee  
Knoxville, TN 37996-1501, USA },
email={barnes@bethe.phy.ornl.gov}
}

\begin{abstract}
This contribution reviews some of the theoretical issues and predictions that
were discussed at HADRON2001. The topics are divided into 
principle areas, 1) exotics, 2) vectors, 3) scalars, and 4) higher-mass states.
The current status of theoretical predictions for each area are summarized,
together with a brief description of experiment. New and detailed 
experimental results are presented 
in the companion Experimental Summary by Klempt. 
\end{abstract}

\date{\today}

\maketitle

\section{Introduction  and  Overview}

\begin{figure}
  \resizebox{12pc}{!}{\includegraphics[height=.3\textheight]{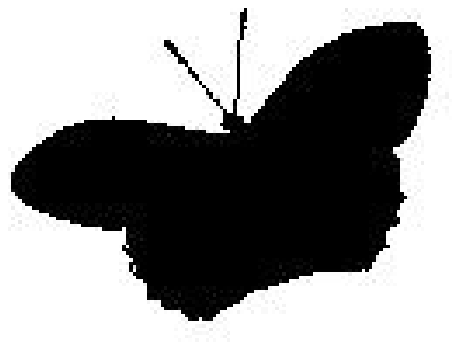}}
\end{figure}

Hadron physics is concerned with the questions of what hadrons exist in nature
and how these hadrons interact and decay.
In each of these areas there are important issues
that are poorly understood. 
Our nominal classification of hadrons as quarkonia,
glueballs and hybrids (and perhaps multiquarks) 
is of course an oversimplification,
and it is not yet clear what resemblance the real hadron spectrum has to our
expectations for gluonics and other exotica.
The most widely used model of open-flavor
hadron strong decays, the $^3$P$_0$ model, is a naive
pair-production prescription with no clear connection to QCD. Finally,
the nature of the strong force between hadrons in general,
which is clearly a very important issue in strong interaction 
physics, remains controversial.

The year 2001 is a transitional period for hadron physics, as was reflected
in the material presented. Two high-statistics experiments using hadron
beams, E852 at BNL and the Crystal Barrel at LEAR, ended several years
ago. Results from several new final states studied at these experiments were
presented here, and some of the results were very interesting indeed;
nontheless it is clear that we are near the end of new results from these
experiments. Hadron spectroscopy using hadron beams will continue
here in Protvino, but will not 
again be a major world enterprise until new facilities
such as GSI and perhaps KEK join this effort. 

In the near future we can expect to see exciting new results from electron beam
and $e^+ e^-$ facilities. For light hadrons this will most noticably 
involve Novosibirsk (with an energy upgrade to an invariant mass of
around 2~GeV) and Frascati (now studying the $\phi$ but with capabilities for
operating at higher mass). These facilities will be complimented by studies
of $c\bar c$ and charm spectroscopy at BES (very nice results for states
above $D\bar D$ threshold were shown here), and in the near future, CLEO-c.
These facilities can also study the very interesting questions in
light meson spectroscopy that can be addressed using two-photon collisions
and initial-state radiation.

Hadron spectroscopy of late has also received contributions from machines such as
LEP and KEK, which were designed for electroweak physics but can make very 
interesting
contributions to light meson spectroscopy, in this case through
two-photon collisions. Experiments that are nominally 
studies of weak interaction physics, such
as charm meson decays, have also rediscovered 
strong interaction physics in the form of important FSIs. 
The implications of these FSIs for light scalar mesons led 
to some interesting interactions between representatives of
the "old" and "new"
cultures in hadron physics in the course of this meeting.

\begin{figure}
  \resizebox{20pc}{!}{\includegraphics[height=.3\textheight]{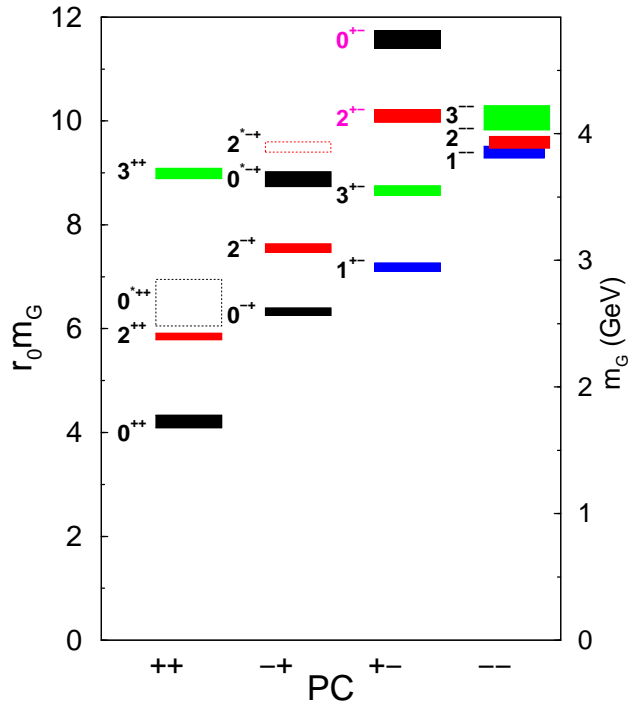}}
  \caption{The quenched LGT glueball spectrum of Morningstar and Peardon 
\cite{CM,LGT_glueballs}.}
\end{figure}

In theory, we also have seen a mix of "old" and "new" 
approaches in this meeting. 
The traditional quark models of hadrons \cite{ESS,YK} remain the most
relevant to experimentalists over the largest part of the $q\bar q$ and
$qqq$ spectrum, since the results are known to be reasonably accurate
numerically, and the radial and orbital excitations of
greatest current interest are readily accessible to these methods.
In parallel, the "first principles" LGT approach 
\cite{CM,TK,PM} has made great progess
in its applications to the spectrum of pure glue and mixed quark-gluon 
states. In the glueball sector the LGT results 
\cite{LGT_glueballs} (Fig.1) are widely regarded
as near definitive (within the quenched approximation), 
which is why we no longer hear suggestions that the
"$\sigma$" or $\eta(1440)$ might be glueballs; LGT has eliminated these
possibilities in favor of a much higher glueball mass scale.
Similarly, the approximate agreement between the predicted LGT scalar glueball
mass and the $f_0(1500)$ has been considered to be a very strong argument 
in favor of a glueball (or mixed glueball-$q\bar q$) 
assignment. Similarly the LGT estimate of the
hybrid mass scale reported at this meeting, which is quite similar to 
the flux-tube model estimate, is considered to be a serious
problem for the light exotic candidate $\pi_1(1400)$. Clearly LGT is
now the leading theoretical approach for estimating the masses of 
gluonic states. Although predictions for the masses of 
the lower-lying excited
mesons and baryons can similarly be extracted from LGT, and in some cases 
should be relatively straightforward since some are the lightest states
in their sector, this important application has not yet received sufficient
attention from LGT groups. The spectrum of excited $q\bar q$ states in LGT 
is obviously a very important topic, which should be considered by LGT
collaborations with improved statistics in future.

The next important step in theoretical technique, both in 
LGT \cite{CM,TK,PM} and
in quark models \cite{EvB}, may be the removal of the "quenched approximation"
through the incorporation of
creation and annihilation of intermediate $q\bar q$ 
pairs. This will lead to several perhaps very important effects, such as 
large mass shifts due to virtual decays. 
The reasons for the success of the naive LGT quenched approximation,
and the closely related quark-model valence approximation,
are important and long-standing questions that can
be addressed in this work.

\eject

\section{Principal Topics}

\subsection{Exotica}

"Exotica" generically refers to states that are not dominantly $q\bar q$ mesons 
or $qqq$ baryons, to the extent that this can be quantified. In this Hilbert
space classification our current expectation is that the possible types of 
exotica are hybrids, glueballs and multiquark systems, with the latter category
including quasinuclear "molecules" and possibly multiquark hadrons.
There
will of course be configuration mixing between these ideal 
"conventional" and "exotica" basis 
states, 
except in the
cases of outright exotic quantum numbers such as I=2 or J$^{\rm PC}=1^{-+}$. 
The amount of configuration mixing will be strongly channel-dependent, and in some
cases may preclude a separation into exotica and conventional hadronic resonances. One
now familiar example is the scalar glueball sector, in which the strong decays
of the 
$f_0(1300)$,   
$f_0(1500)$ and 
$f_0(1710)$ are all far from expectations for pure $q\bar q$ or glue states,
due perhaps to very large 
$|n\bar n\rangle 
\leftrightarrow 
|G \rangle 
\leftrightarrow 
|s\bar s\rangle $
mixing effects.
Alternatively, in the cases of exotic flavor or J$^{\rm PC}$ we can be certain that
identification of a resonance is an indication of a state beyond the naive
quark model of $q\bar q$ mesons and $qqq$ baryons. The identification of the
spectrum of such states is the most important task for QCD spectroscopy
at present.

Theorists derived the expected spectrum of hybrids (including
J$^{\rm PC}$-exotics) in various models beginning in the mid
1970s. It is now widely accepted that hybrid mesons span all J$^{\rm PC}$, and
the lightest hybrid exotic should be a $1^{-+}$. In some models such
as the flux tube model there are additional exotics present in the lowest multiplet,
specifically $0^{+-}$ and $2^{+-}$. These states are also expected in 
the bag model, but at rather higher mass.) The search for such exotic quantum
numbers was given a strong incentive by the flux tube calculations of Isgur, Kokoski
and Paton\cite{IKP}, who predicted very characteristic decay modes for
hybrids, specifically S+P final states such as $f_1\pi$ and $b_1\pi$.
Their mass estimate of ca. 1.9~GeV was somewhat higher than was predicted
earlier, for example using the bag model. The restricted S+P decay modes 
compensated for the increased phase space at the higher flux-tube mass scale, 
so the flux-tube decay 
calculations found that some
hybrids, notably a $\pi_1(1900)$, should be relatively narrow.
Of the other relatively narrow states predicted by this model,
the most remarkable are an "extra" $\omega$ that would favor $K_1K$ modes
and an "extra" $\pi_2$ that would decay strongly to $b_1\pi$.
(The $b_1\pi$ mode is forbidden to the quark model $\pi_2(1670)$ because the
$^1$D$_2 \to  ^1$P$_1 + ^1$S$_0$ transition is spin singlet
to spin singlet, which vanishes in the $^3$P$_0$ decay model.)

Relatively recent theoretical results on the hybrid mass scale
in LGT were presented at this meeting. 
These results are more accurate at
higher quark masses, due to the use of a nonrelativistic 
expansion of the QCD action; this leads to mass predictions 
that have much smaller statistical errors for states that
incorporate heavy quarks.
The masses predicted for the $1^{-+}$
$b\bar b$-
and
$c\bar c$-hybrids in the most recent calculations (reported here
by Morningstar \cite{CM}) are
$M_{b\bar b \ hybrid} \approx 10.9$-$11.0$~GeV
and
$M_{c\bar c \ hybrid} \approx 4.3$~GeV, which should be very useful as 
motivation for future studies of the higher-mass $c\bar c$ system 
at CLEO and BES. (Models typically anticipate approximately degenerate 
$1^{-+}$ and $1^{--}$ hybrids, so we expect to see an "extra"
$c\bar c$ $1^{--}$ in $e^+e^-$ annihilation at about this mass.)
The especially interesting $n\bar n$-hybrid with $1^{-+}$ quantum 
numbers is predicted to lie at about 1.9-2.1~GeV \cite{CM}, 
quite close to 
the flux tube model estimate. As a final interesting point, NRQCD is
now finding results for the masses of nonexotic hybrids as well; 
a level ordering of
$2^{-+} > 1^{--} > 1^{-+} > 0^{-+}$ found by Drummond {\it et al} \cite{Drummond}
using NRQCD LGT was reported at this meeting
\cite{CM}; this ordering was 
predicted by the bag model. In contrast
the usual flux tube model results predict
these states to be
degenerate. This may be another area in which LGT can act as 
{\it de facto} theoretical QCD 
data that can be used to distinguish between different
intuitive models, pending experimental results.

Regarding the {\it de jure} data on exotics,
two candidate J$^{\rm PC}$ exotic meson resonances have been proposed, 
both with
I=1, J$^{\rm PC}=1^{-+}$ quantum numbers; the 
$\pi_1(1400)$ and $\pi_1(1600)$. Obviously, establishing (or refuting) 
these candidate exotic resonances is of paramount importance for the future
development of spectroscopy, since if confirmed they provide a benchmark
for the mass of the lightest exotic resonance and the energy scale of
exotic radial excitations. Unfortunately the $\pi_1(1400)$ signal (in
$\eta\pi$) is rather weak, so it is difficult to distinguish this resonance
interpretation
from a nonresonant background phase. (This simple statement
summarizes two decades of experiment.)

At this meeting we have heard from the VES collaboration \cite{VD} that they 
now have no clear preference for a 
$\pi_1(1400)$ resonance interpretation; they find fits of
similar quality from a nonresonant signal. Since the favored theoretical methods
anticipate a
much higher mass of ca. 1.9-2.0~GeV for the lightest hybrid meson 
multiplet, which includes the lightest expected I=1 $1^{-+}$ exotic, theorists
would generally be happier if the $\pi_1(1400)$ were to be reinterpreted
as a nonresonant signal, and the very clearly resonant $\pi_1(1600)$ were
to replace it as the lightest exotic. Of course we must be cautious 
here because
these predictions are for an unfamiliar system in the quenched approximation;
the mass shifts due to couplings to virtual meson loops are currently unclear,
and may be rather large. This will be 
a very important issue for future theoretical studies.

\begin{figure}
  \resizebox{20pc}{!}{\includegraphics[height=.3\textheight]{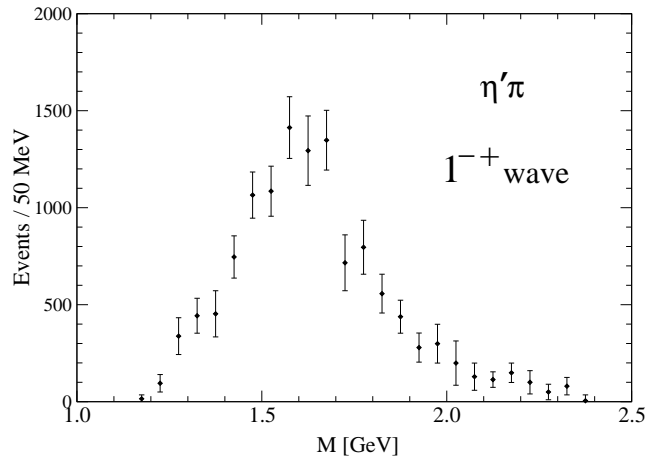}}
  \caption{The E852 $1^{-+}$ wave in $\eta'\pi^-$, showing a dominant $\pi_1(1600)$ exotic
\cite{Popov}.}
\end{figure}

In contrast, the $\pi_1(1600)$, which is already claimed in $\eta'\pi$, $\rho\pi$ and
$b_1\pi$ final states, may now be clearer. In their contribution \cite{Popov}, E852
showed results from the $\eta'\pi$ final state, in which the dominant low-energy
resonance is the $\pi_1(1600)$ (see Fig.2). The usually dominant $a_2(1320)$ is
much weaker in this channel due to small branching fraction of
$B(a_2\to \eta'\pi)\approx 0.5 \% $; this leaves a remarkably robust $1^{-+}$ 
exotic
wave, which if confirmed as resonant (see \cite{VD} for a cautionary note)
will presumably be a benchmark for future studies of exotics.
Note that the fitted width of $\Gamma_{tot}(\pi_1(1600)) = 340\pm 40 \pm 50$~MeV
is rather broader than the earlier estimates from the $\rho\pi$ final state.

\subsection{Vectors}

The conference began with a summary by Donnachie of the status of light vectors
\cite{AD}.
Although this might appear to be a rather specialized topic, in my opinion it merits
a special section because much of the future work on light meson spectroscopy will
concentrate on the vector sector. This is because the new and upgraded 
$e^+e^-$ machines at Frascati and Novosibirsk produce vector mesons 
in $e^+e^-$ annihilation, and future photoproduction facilities 
such as HallD at Jefferson Lab will also produce $1^{--}$ states (not uniquely,
but vectors should also dominate diffractive photoproduction).

This limitation to $1^{--}$ states is an advantage in
disguise; as usual in the 1-2 GeV mass region we have broad overlapping resonances,
but since only $1^{--}$ is important in 
$e^+e^-$ annihilation, we expect to produce only a few resonances per flavor channel. 
Thus it
should
be possible to establish clearly what states are present in the 
light meson spectrum, and whether there is indeed an overpopulation of states
relative to the naive
$q\bar q$ quark model. 

Application of 
the quark potential model to the $n\bar n$ sectors leads to predictions of 
2$^3$S$_1$ 
radial excitations near 1.5~GeV, L=2
$^3$D$_1$ $n\bar n$ states near 1.7~GeV, and a 3S radial excitation near
2.1~GeV. 
Experiment appears to support the existence of these 2S and D states (Fig.3),
with  $\rho$ and $\omega$ flavor states roughly degenerate, and some evidence for
$K$ and $\phi$ analogues expected about 0.12 and 0.25 GeV higher in mass. 
Note however that
$K^*(1410)$ appears surprisingly light if it is a partner to 
$\rho(1465)$ and $\omega(1420)$ 2S states. 
(A parenthetical note: Could this
indicate the presence of the $1^{-+}$ exotic, with 
$1^{-+}$-$1^{--}$ mixing in the kaon sector analogous to the $K_1$ states?)

Of course only the $\rho^\circ, \omega$ and  $\phi$ are accessible to
$e^+e^-$, and again we are fortunate in $e^+e^-$ because the relative flavor cross section
ratios for $\rho^\circ :\omega : \phi $ of $9:1:2$ are known. 
(Some additional suppression of
$s\bar s$ production is expected, due to the
larger $m_s$.) 
With only two $q\bar q$ states anticipated by theorists per flavor sector 
between $\approx$ 1.5~GeV and $\approx$ 2.0~GeV, this
problem sounds almost too simple!

There are two complications that have left the vector sector in a confused state
despite decades of previous study, primarily using $e^+e^-$ and photoproduction facilities.
The first and most important problem is that the more accessible $\rho^\circ$ and $\omega$
states are quite broad, so we face the famous problem of overlapping
resonances. Another difficulty is that we anticipate a $1^{--}$
hybrid meson multiplet somewhere in this mass region (degenerate with the
$\pi_1$, in the flux tube model), so we may have not two but three states
(2S, D, H) in this mass region. Actually this is again fortunate, since it affords us
the opportunity to study conventional $q\bar q$ and hybrid states in a very
restricted slice through Hilbert space, in a channel in which the important mixing effects
can also be investigated. 
What we learn from excited vectors as an isolated case study may be
crucial in helping us to understand the other nonexotic 
sectors of light meson spectroscopy.

\begin{figure}
  \resizebox{22pc}{!}{\includegraphics[height=.3\textheight]{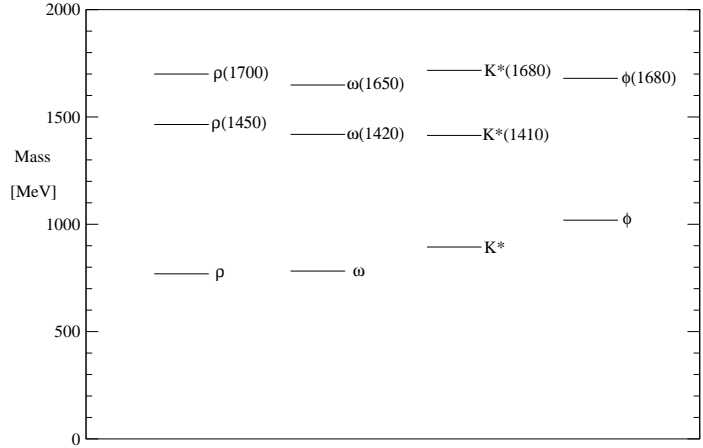}}
  \caption{Experimental vector mesons below 2 GeV.}
\end{figure}

In addition to the location of the individual levels, which may
well be quite different from quark model expectations if $q\bar q \leftrightarrow$
hybrid mixing is important, the strong decay modes of the vectors will be 
especially interesting. This is because much of our theoretical
"scaffolding" for hadrons and their strong decays relies on the so-called
$^3$P$_0$ model, which assumes that strong decays take place through 
production of an additional $q\bar q$ pair with vacuum quantum numbers
(J$^{\rm PC}=0^{++}$, hence $^3$P$_0$). 
Other popular decay models such as the flux tube 
decay model are relatively minor variants of the original $^3$P$_0$ model,
introducing for example a smooth spatial modulation of the pair production
amplitudes.
\begin{table}
\caption{
Theoretical partial widths of 2S, 1D and hybrid $\rho$ states.}
\label{tabrho}
\begin{tabular}{lccccccccc}
 & $\pi\pi$ & $\omega\pi$  & $\rho\eta$ & $\rho\rho$ & KK & K$^*$K
& $h_1\pi$ & $a_1\pi$  & total\\
\hline
$\rho_{2S}(1465)$ & 74. & 122. & 25. & - & 35. & 19. & 1. & 3. & 279.  \\
$\rho_{1D}(1700)$ & 48. & 35. & 16. & 14. & 36. & 26. & 124. & 134. & 435.  \\
$\rho_H(1500)$ & 0 & 5 & 1 & 0 & 0 & 0 & 0 & 140 & $\approx 150$
\end{tabular}
\end{table}
Although this model has been employed by theorists to 
reach a broad range of conclusions about hadrons 
(such as S+P decay modes for hybrids, and a list of "missing baryons" which
are purportedly missing because they
couple weakly to $\pi N$), it has been tested in disturbingly few decays.
The most sensitive and well known tests are in decays of axial vectors to
vector plus pseudoscalar. This channel allows both S- and D-waves in the 
VPs final
state, so one can determine the relative magnitude {\it and} sign of S and D
through the decay product angular distribution. The D/S ratio is quite sensitive
to the quantum numbers of the  $q\bar q$ pair produced in the decay, and the
observed value of $\approx +0.28$  
in $b_1\to \omega \pi$ \cite{Popov} 
strongly supports the $^3$P$_0$ model.
The decay $a_1\to \rho\pi$ is predicted to have a D/S ratio of -1/2 times the
$b_1\to \omega \pi$ ratio, which is also reasonably well satisfied.
(Actually the D/S amplitude ratio is complex, since S- and D-wave 
VPs final states develop
different FSI phases. This allows one to determine
the phase shift difference $\delta_S - \delta_D$ in the $\omega\pi$
system at the $b_1$ mass, which has only recently been 
appreciated and exploited \cite{NN}.)
These two measurements, and some additional support from other axial vector 
decays in kaon and charm systems, are the only clear checks of this very widely
used decay model.

When applied to these excited vector states, these strong decay models 
predict markedly different favored modes 
for the different states \cite{DK,BCPS,CP}, 
that may be useful as signatures and to establish mixing angles
between different vector basis states.
In Table 1 we show results for the three $\rho$-type
excited vectors; evidently these have comparable theoretical widths but
very distinct branching fractions. The broad $4\pi$ states from $h_1\pi$
and $a_1\pi$ are predicted to arise from the 2D (comparable $h_1\pi$ and
$a_1\pi$) and H ($a_1\pi$ only), whereas 2S should couple strongly to neither,
instead populating $\pi\pi$ and $\omega\pi$. If this is accurate, it shows
the importance of measuring as many final states as possible, especially 
since mixing of these basis states may be important. 

As noted by Donnachie, the existing data has many gaps in energy and final state
coverage, but it is clear that the $4\pi$ modes do not appear to agree
with Table~1. There is evidence for $a_1\pi$ dominance of broad $4\pi$ states
from the ratio of $\pi^+\pi^-\pi^\circ\pi^\circ / 2\pi^+ 2\pi^-$, which would
only be expected from a hybrid! (This assuming the flux-tube model of
hybrid decays is accurate.) With accurate measurements of the final states
in Table~1, we should be able to distinguish the $\rho$ excitations present
in this channel, and should learn about state mixing and strong decay amplitudes
in the process. 

Table~1 lists only $\rho$ states. Donnachie noted that
the flux tube decay model predicts a very narrow $\omega_H$ $1^{--}$ hybrid,
coupled strongly only to $K_1K$ decay modes \cite{CP}. 
If this state is near the $\pi_1(1600)$ mass these modes are closed,
and the flux-tube suppressed mode of 
$\rho\pi$ is expected to lead to a total width of only 
$\sim 20$~MeV \cite{AD,CP}. This remarkable prediction
strongly motivates a simultaneous study of $\omega$-flavor $1^{--}$ states.

It may be that the  $^3$P$_0$ model is inaccurate outside
the $1^+$ channel, in which case most of our predictions 
of hadron strong decays will
be inaccurate. Evidence for a failure of the 
$^3$P$_0$ model in $\pi_2\to\rho\omega$
was presented by E852 at this meeting, 
which I will mention in the section on higher-mass
states.

\subsection{Scalars}

\subsubsection{Introduction}

I will first discuss the famous "980 states", in which there has been clear progress
recently, and a close interplay between theory and experiment may have clarified much
about the nature of these states. 
These results were clearly considered by many to be
the most interesting 
presented at this meeting. 
Next I will briefly discuss the broad 
"$\sigma$" scalar and its purported strange partner, which were discussed at this 
meeting at some length 
but (as usual) no clear concensus as to the best description
of the physics was evident. Finally I will suggest interesting future possibilities for 
clarifying the nature of the various scalars in the next round
of experiments. Although the scalar sector includes the scalar glueball, and allows
one to address the very important question of glueball-quarkonium mixing, little new 
experimental material was presented at this meeting, so I will not discuss 
glueballs as a separate topic.

\subsubsection{"980" States}

The two mesons near 980~MeV, once the S$^*$ and $\delta$, now the $f_0(980)$ and
$a_0(980)$, have long attracted attention as being anomalous in many of
their properties. Although close to degenerate, so that we might expect them to be
nonstrange $n\bar n$ I=0,1 partners, their very strong coupling to
$K\bar K$ suggests that these are actually not conventional $n\bar n$ quark model states.
Other problems are that 
their strong total widths are much smaller than expectations for 
$n\bar n$ at this mass, their masses are well below those of other P-wave 
$n\bar n$ states and are just below the $K\bar K$ threshold, 
and their electromagnetic
couplings 
(specifically $\gamma\gamma$) are much weaker than we would expect for $n\bar n$.
This list of problems can be expanded considerably. 

Historically three models of these states have been 
considered by theorists. These suggest 
that the 
$f_0(980)$ and $a_0(980)$ 
might be four-quark clusters
(primarily supported by Achasov {\it et al.}),
weakly-bound kaon-antikaon quasinuclear states
(Weinstein and Isgur),
or simply $q\bar q$ quark model states, whose properties happen to 
differ from our
naive expectations for ordinary mesons. Of course all accessible
basis states will mix in physical hadrons, perhaps significantly, 
so we should more properly regard these models as suggestions regarding
which component dominates in the expansion 
\begin{equation}
|980\rangle 
=
c_{q\bar q} |q\bar q\; \rangle + 
c_{q^2\bar q^2} |q^2\bar q^2 \; \rangle + 
c_{K\bar K} |K\bar K\; \rangle + 
\dots \ .
\end{equation}
Of course the coefficients are actually spatial wavefunctions, so the distinction
between $|q^2\bar q^2 \; \rangle $ and $|K\bar K\; \rangle $ basis states is 
rather qualitative.

An important test proposed to distinguish between these descriptions
(assuming dominance
of one basis state) arises in $\phi(1020)$ radiative decays. 
In both the four-quark and
$K\bar K$-molecule models it is assumed that the 980 states are produced in
$\phi$ radiative transitions by photon emission from 
a virtual $K\bar K$ loop, 
with a direct photon coupling to the $K^+K^-$ loop
but not to $K^\circ\bar K^\circ$.
The corresponding decay rate was evaluated by Achasov, Devyanin and Sheshtakov 
\cite{ADS} and by Close, Isgur and Kumano \cite{CIK}. Their result for 
the branching fractions is
\begin{equation}
B(\phi \to \gamma f_0(980))
=
B(\phi \to \gamma a_0(980))
\approx
(2.0\pm 0.5) \cdot 10^{-4} \cdot F(R)^2\ ,
\end{equation}
where $F(R)$ is a form factor that depends on the spatial wavefunctions
of the mesons; $F(R)$ would be  
unity for a pointlike $K^+ K^- m(980)$ coupling.  
In contrast, a $q\bar q$ picture of the 
$f_0(980)$ and $a_0(980)$ 
would predict very small branching fractions 
of perhaps $10^{-6}$ if $f_0(980)= s\bar s$, and even smaller were they
$n\bar n$.

Klempt will discuss the experimental results 
for these branching fractions
from Novosibirsk and Frascati 
in his experimental summary.
Here I will simple note that they are comparable in scale to the $\approx 1$-$3 \cdot 10^{-4}$
quoted above, but the weaker result that the branching fractions are equal,
\begin{equation}
{B(\phi \to \gamma f_0(980))
\over
B(\phi \to \gamma a_0(980))
}\bigg|_{theory} = 1
\end{equation}
is not at all well satisfied! Instead the experimental ratio is
\begin{equation}
{B(\phi \to \gamma f_0(980))
\over
B(\phi \to \gamma a_0(980))
}\bigg|_{expt.} 
\approx 4 \ .
\end{equation}
If we reconsider the charged-kaon-loop radiative decay models to see what might have 
gone wrong, we find that the ratio of unity follows from the assumption that both
980 states 
are isospin eigenstates. It was instead argued long ago in both 
$q^2\bar q^2$ \cite{ADS} and $K\bar K$ \cite{TB_980s} models that one should anticipate
important
isospin violation in these states. 
For $q^2\bar q^2$ this arises from mixing through
nondegenerate $K^+K^-$ and $K^\circ\bar K^\circ$ loops, and for $K\bar K$ from the
fact that these are weakly bound $K\bar K$ systems, with 
zeroth-order $K^+K^-$ and $K^\circ\bar K^\circ$ masses
that differ
by an amount comparable to the binding energy. There was already evidence
for isospin mixing in these states, through $\pi\pi\to \pi\eta$ transitions
evident in E852 data, and through evidence for central production of both the
$f_0(980)$
and
$a_0(980)$. 
The central production data suggests a mixing angle near $15^\circ$, which led Close and
Kirk \cite{C2001} 
to a modified prediction for the radiative transition ratio of
\begin{equation}
{B(\phi \to \gamma f_0(980))
\over
B(\phi \to \gamma a_0(980))
}\bigg|_{theory} = 3.2\pm 0.8 \ ,
\end{equation}
which is consistent with observation. The absolute scale of the rates suggests 
a hard form factor $F(R)\approx 1$, which supports the picture of a compact
four-quark system. Close and Kirk interpret this as evidence for a combination of
a $K\bar K$ system with a compact $q^2\bar q^2$ core. 

In summary, we have clear and consistent evidence of
a large isospin mixing angle in these states 
from three experimental processes,
at a level not seen in other hadrons. This
is a very interesting result indeed. A future calculation that is immediately
suggested by this observation is to determine the mixing angles predicted
by the two models of isospin violation, mixing through kaon loops versus
mixing due to weak binding of nondegenerate $K^+K^-$ and $K^\circ\bar K^\circ$ systems.

This evidence of a large isospin mixing angle between the nominally
I=0 $f_0(980)$ and I=1 $a_0(980)$ immediately suggests several interesting
measurements, which might check this result and independently 
determine the mixing angle.
These include 1) the $\gamma\gamma$ widths, which were also predicted to be equal
for both states
because of photon coupling to the charged kaon loop alone, and which we 
therefore expect to be skewed in favor of the $f_0(980)$ by the same ratio
as the radiative transition; 2) the relative annihilation decay rates of 
$J/\psi \to \phi(\pi\pi)$ and  $J/\psi \to \phi(\pi\eta)$ 
(with isospin eigenstates we would
expect to see no 980 signal in $\pi\eta$, since this is driven by an
$s\bar s$ source; similarly for $D_s \to \pi (\pi\pi)$ and 
$D_s$ to $\pi (\pi\eta )$). Finally, radiative transitions such as 
$a_0(980)\to \gamma\omega$ and
$f_0(980)\to \gamma\omega$ 
can be used to quantify the $n\bar n$ components in the 980 
states, since E1 radiative transition amplitudes of light quarkonia 
are reliably calculable in the quark model.

\subsubsection{Broad Scalars ("Let Sleeping Dragons Lie.")}

Discussions of the status of broad scalars have appropriately spanned decades.
The contending "camps" in this area have long since settled on favorite
explanations of the low-energy "$\sigma$" and "$\kappa$" effects, and these
views are held with the tenacity of religious convictions. This situation makes for
bad science, and we may need new, independent experimental information about the light
scalar sector before we can make any progress in our understanding of broad
scalar states.

At this meeting we have heard discussions of the relatively recent information
on the light $\pi\pi$ and $K\pi$ systems that has come from charm decay experiments.
In these experiments it was noted that there are clear low-energy enhancements in 
I=0 $\pi\pi$ and I=1/2 $K\pi$ subsystems, which can be fitted by {\it very} light
scalar resonances. Specifically, masses of $\approx 480$~MeV and $\approx 800$~MeV
were quoted for 
"$\sigma$" and "$\kappa$" states \cite{Goebel}. 
This is probably a premature conclusion, since 
only the low-energy tails of the purported resonance phase shifts are actually in evidence 
in the charm data; the crucial observation of a complete Breit-Wigner phase 
motion through 180$^\circ$ has not been made. It was noted here by
Ochs \cite{Ochs} and by Pennington that the elastic $\pi\pi$ and $K\pi$ phase 
shifts themselves do not show evidence of "complete" low mass scalar resonances,
so concluding that these exist based on the charm decay data in isolation,
which only covers part of the range of invariant mass that has already been studied
in light hadronic processes, is unjustified.
The discussions at HADRON2001 following the charm decay presentations 
suggested 
that the charm decay analyses should include
what is already known about these phase shifts over the full relevant mass
range, for example through the parametrization of Au, Morgan and Pennington
\cite{AMP}. 

Experience suggests that progress may follow from 
a high-statistics study of a new production mechanism in the relevant mass region,
as was provided by $\phi$ radiative decays for the 980 states. I would suggest that 
future high-statistics two-photon
collisions, especially $\gamma\gamma\to\pi^\circ\pi^\circ$, may be definitive in
resolving the resonances present in the light I=0 scalar channel. This reaction
is quite simple (only S- and D-waves are produced significantly at low energies),
and with high statistics it should be possible to determine the S-wave phase motion
through interference with the $f_2(1270)$ D-wave. This reaction was studied 
earlier by the Crystal Ball
collaboration \cite{CBal}, 
{\it albeit} with quite limited statistics; their results showed
a broad scalar signal under the $f_2(1270)$, but the data was not adequate for a 
determination of the mass and width. If one could track the phase motion
of the S-wave in this process 
(perhaps augmented by $\gamma\gamma\to\pi^+\pi^-$ and $\gamma\gamma\to\eta\eta$
data) it should be possible to identify the lighter $f_0$ scalar
resonances. We should be aware that slowly-varying background phases are also
present, which may significantly modify the fitted resonance parameters in this channel;
as an example, the J\"ulich group note that $t$-channel $\rho$ exchange in 
$\pi\pi$ scattering with a realistic $\rho\pi\pi$ coupling strength can explain most
of the low-energy $\pi\pi$ phase shifts in both I=0 and I=2 channels 
\cite{Juelich}. 
Thus we may not learn where the light scalar resonances lie until we have understood
nonresonant "background" phase shifts as well.

LGT predictions for scalar 
$q\bar q$ masses
would also be of great interest.
Although these would be "quenched" results, these bare numbers
actually are used in some models of $\pi\pi$ scattering, and in any case there is 
so much uncertainty in this field at present 
that {\it any} more definitive theoretical result would
be important. Just as the large LGT glueball mass scale in Fig.1
\cite{LGT_glueballs} has eliminated the 
"$\sigma$" and 
$\eta(1440)$ from serious contention as glueball candidates, so LGT results
for the scalar $q\bar q$ spectrum could help to identify the more plausible 
scenarios in this most obscure and controversial sector of Hilbert space.

\subsection{Higher-mass States}

\subsubsection{Heavy Quarkonium}

We heard several interesting experimental contributions about heavy quarkonium
at HADRON2001,
specifically about the charmonium system. 
Although little new theoretical activity
was reported in this field (the exception is heavy-quark hybrid masses from LGT),
we will presumably see future theoretical interest in the charmonium system 
in response to high statistics studies at BES and
CLEO-c. For this reason it seems appropriate to at least mention 
some of the charmonium results reported, and to suggest some
possibly interesting 
questions for future experimental and theoretical
investigation.

\begin{figure}
  \resizebox{20pc}{!}{\includegraphics[height=.3\textheight]{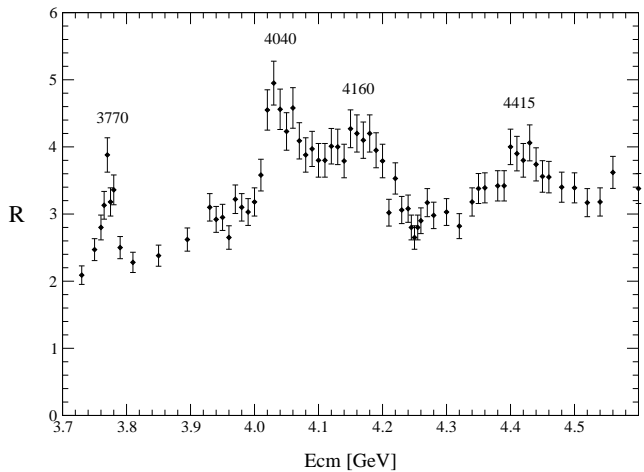}}
  \caption{The BES measurement of R \cite{WLi}.}
\end{figure}

First, BES has reported results for the inclusive hadron cross section ratio $R$ in
the region above open charm threshold \cite{WLi} (Fig.4).
This is an important advance,
as the rather noisy previous results from the late 1970s suggested the higher-mass
resonances $\psi(3770), \psi(4040), \psi(4160)$ and $\psi(4415)$ but were far from definitive. 
Only 
these four $c\bar c$ resonances are
regarded as established
above open charm threshold, and their masses are consistent with 
potential model expectations for 
1$^3$D$_1$,
3$^3$S$_1$,
2$^3$D$_1$
and
4$^3$S$_1$ 
levels (in order of increasing mass).

Despite this agreement of masses, there are serious problems with the preperties
reported for these states relative to potential model expectations. The 
$\psi(3770)$
and
$\psi(4160)$
should both appear quite weakly in $e^+e^-$ if they are D-wave $c\bar c$ states,
since the wavefunction at contact vanishes in this case. Instead the $e^+e^-$
width of the $\psi(3770)$ is much larger than expected, and the reported
$\psi(4160)$ $e^+e^-$ width is comparable to the 
nominally 3$^3$S$_1$ $\psi(4040)$.
Of course this is based on the old, rather noisy, measurements. The $\psi(4160)$
signal in the new BES data appears weaker, and when fitted this new $e^+e^-$
width may be rather 
smaller than previous estimates. 

The exclusive strong branching fractions of these higher-mass 
$c\bar c$ states will also be 
very important measurements. The existing claims for strong branching fractions
include an estimate that the $\psi(4040)$ favors the 
$D^*\bar D^*$ 
mode over
$D\bar D$ by about a factor of $\sim 500$ \cite{KS,SFT}, 
despite the absence of $D^*\bar D^*$ phase space!
(Recall $M(D^*) = 2.01$~GeV.) This remarkable result previously 
led to suggestions that the
$\psi(4040)$ might be a $D^*\bar D^*$ molecule. The conventional $c\bar c$
description nonetheless appears plausible, 
in view of the agreement with the predicted
mass of the 
3$^3$S$_1$
$c\bar c$ level. The $\psi(4040)$
$e^+e^-$ width, which is comparable to the 
$e^+e^-$ widths
of the $\psi(3686)$ and $\psi(4415)$
2S and 4S radial excitations, also supports a $c\bar c$ $\psi(4040)$ assignment.
The unusual strong branching fractions may be due to
nodes in the strong decay amplitudes; the nodes in the
3S radial wavefunction may well have produced 
counterintuitive branching fractions for the $\psi(4040)$.
Clearly, reasonably accurate measurement of the exclusive 
branching fractions of 
the higher $c\bar c$ states to all open charm final states will be an extremely
interesting set of measurements, which 
can be used as detailed tests of 
strong decay models.   

One 
limitation of $e^+e^-\to\gamma\to q\bar q$ 
annihilation is that it produces only $1^{--}$ states.
One can extend these studies to the two-photon collision process
$e^+e^- \to e^+e^- \gamma \gamma$, $\gamma \gamma \to q\bar q$, to search for
states with even C-parity. These two-photon widths are intrinsically interesting
to theorists, since they can be calculated in quark models, and may 
provide sensitive tests of the quark model states. 
Fig.4 shows a new measurement of a candidate $a_2(1750)$ radial excitation
in $\gamma\gamma$, reported here by the BELLE Collaboration \cite{SHou}. 
The relative
two-photon partial widths of a given J$^{\rm PC}$ flavor 
multiplet vary with flavor as $f: a: f' = $ 
$25: 9: 2$, so two-photon couplings can be used 
to identify flavor partners of 
a given state, or quantify the level of flavor mixing. (There is some
suppression of the heavier $s\bar s$-$\gamma\gamma$ coupling.) 
Two-photon couplings may also be useful in distinguishing different
types of scalar states, since we naively expect glueballs and multiquark states to have rather 
smaller $\gamma\gamma$ couplings than $n\bar n$ states. 
In contrast, in the quark model a light scalar 
$f_0^{(n\bar n)}(1300)$ is predicted to have a two-photon width of $\approx 5$~KeV, larger than
any other light $n\bar n$ meson. 

\begin{figure}
  \resizebox{22pc}{!}{\includegraphics[height=.3\textheight]{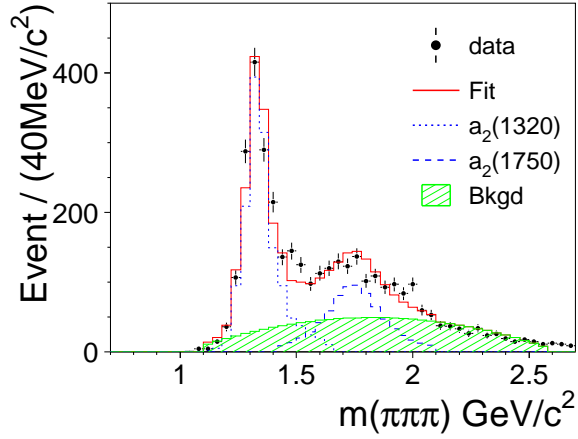}}
  \caption{An example of $\gamma\gamma$ production of a higher-mass $q\bar q$ state,
from Belle \cite{SHou}.}
\end{figure}

Two-photon couplings of charmonia are very interesting in part because they allow us to test
calculations of $q\bar q\to \gamma\gamma$ widths in a regime in which the
nonrelativistic quark model should give reasonably accurate results. 
Typical theoretical predictions are 
$\approx 5$-$7$~KeV 
for the $\eta_c(2980)$ and
$\approx 0.5$-$2$~KeV  for the P-wave $c\bar c$ states $\chi_0$ and $\chi_2$. The ratio of
$\chi_0 / \chi_2$ partial widths varies over the range $\approx 3$-$10$, depending on 
theoretical assumptions.
These measurements 
of $\gamma\gamma$ charmonium widths have a long history of uncertainty, due to 
the intrinsically small $O(\alpha^4)$ cross sections.
It is now clear that the experimental $\eta_c(2980)$ $\gamma\gamma$ width 
\cite{AT} is not far from
theoretical expectations. The P-wave states have somewhat smaller $\gamma\gamma$ widths
and less characteristic decays, and so have been more difficult to measure. 
One competing technique that
proved quite successful was to use $p\bar p$ annihilation to make the $c\bar c$ state,
followed by detection of $\gamma\gamma$ against a very large hadronic background.
(This was done by E760 and E835 at Fermilab.) 
A new BELLE measurement of the $\gamma\gamma$
width of the tensor $\chi_2(3556)$ in $e^+e^-$ collisions was
reported here
\cite{SHou}, 
\begin{equation}
\Gamma_{\gamma \gamma} (\chi_2 )\bigg|_{\rm BELLE} 
= 0.84(0.08)(0.07)(0.07)\ {\rm KeV}\ 
\end{equation}
which is about a factor of three larger than the Fermilab result
\begin{equation}
\Gamma_{\gamma \gamma} (\chi_2 )\bigg|_{\rm E835} 
= 0.270(0.049)(0.033)\ {\rm KeV}\ ,
\end{equation}
presented here by Tomaradze \cite{AT}.
This is about a $4\sigma$ difference, so the discrepancy does 
appear significant. I am amused to note that a 
previous $\chi_2(3556)\to\gamma\gamma$ calculation 
\cite{TB_chi2} found a value of 
$\Gamma_{\gamma\gamma}(\chi_2) \approx 0.56$~KeV, comfortably between the
two experimental results.

\subsubsection{A Striking $^3$P$_0$ Decay Model Failure}

One especially interesting new result reported at this meeting concerned
resonances observed in the $\rho\omega$ final state. This is
very important theoretically because the $VV$ system can have $S=0,1$ and $2$,
so there is considerable scope for testing strong decay models. (Recall
that we have all been using the $^3$P$_0$ model or variants to predict
light meson decays, $D$ meson decays, hybrid decays, missing baryons and so
forth for decades, but this model has seen little in the way of sensitive
tests of the quantum numbers of the $q\bar q$ pair formed in the decay.) 
The historically convincing 
angular correlation tests were in $1^+$ decays to
VPs final states, specifically $b_1\to\omega\pi$ and $a_1\to\rho\pi$, in which
both S- and D-wave VPs final states are produced. The model does predict these 
two D/S 
ratios approximately correctly, but it has seen few sensitive tests in
other $J^P$ sectors. When applied to decays into $VV$ final states, the model typically
predicts a nontrivial pattern of large, small or identically zero 
decay amplitudes, which can be compared to these new results
on $\rho\omega$.

The $\pi_2(1670)$ is an interesting initial state for these decay model
tests; it is a spin singlet (${}^1$D$_2$ in the quark model),
so many decay amplitudes are predicted to be zero due vanishing spin matrix
elements. For example, the decay $\pi_2\to b_1\pi$ is strictly 
forbidden in the
$^3$P$_0$ model, since this would be an S=0 to S=0 
transition (the mesons all have S=0);
the 
$^3$P$_0$ 
transition operator has S=1 
$(\vec \sigma \cdot \vec p)$, so there is no 
S=0 to S=0 matrix element. The fact that this branching
fraction is indeed quite small is one of the few recent decay model tests.

On considering the decay $\pi_2\to\rho\omega$, one
immediately finds a dramatic failure, assuming that the newly 
reported experimental decay amplitudes are correct.
The $2^{-+}$ $\rho\omega$ system can in general have the quantum numbers
$^3$P$_2$, 
$^3$F$_2$, 
$^5$P$_2$,
and 
$^5$F$_2$, but
the $^5$P$_2$ and $^5$F$_2$ $\rho\omega$ final states are forbidden
to $\pi_2\to\rho\omega$ in the $^3$P$_0$ model, since we have an S=0 
initial state and an S=1 transition operator. We should only find 
$^3$P$_2$ and $^3$F$_2$ $\rho\omega$ final states. Of these 
we expect the $^3$P$_2$ $\rho\omega$ wave
to dominate $\pi_2\to\rho\omega$, since there is little phase space.

\begin{figure}
  \resizebox{20pc}{!}{\includegraphics[height=.3\textheight]{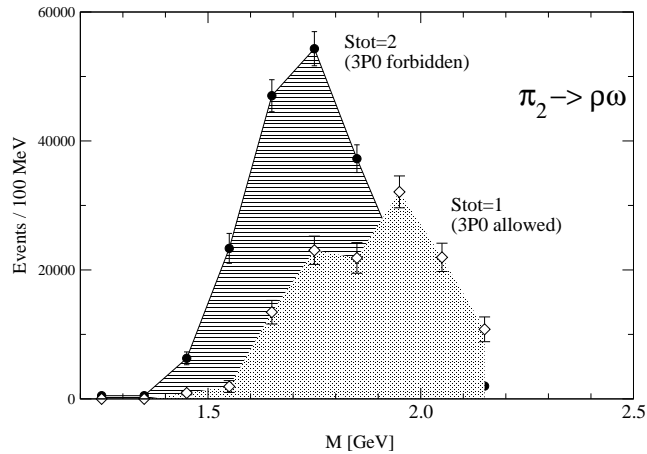}}
  \caption{Final S$_{tot}$ amplitudes in $\pi_2\to\rho\omega$, showing violation of $^3$P$_0$ model expectations
\cite{Popov}.}
\end{figure}

Experimentally only the S=2 $\rho\omega$ final state is observed to
peak in the $\pi_2(1670)$ region, which implies that this decay is
dominated by a {\it spin tensor} transition. This final state 
might be
generated by $q\bar q$ pair production from a transverse gluon, but it
is certainly not anticipated by the usual $^3$P$_0$ strong decay model.
Subsequent angular analysis of the $VV$ system may provide other interesting
results regarding the mechanism of these still poorly understood strong 
decays. 

\section{Summary}

In this report I have briefly summarized several interesting topics
that were discussed in presentations at HADRON2001. These included
evidence of and expectations for exotic mesons, the status of light 
vector mesons, the very interesting new results on the $980$ states, 
new results for $R$ in the open-charm region, and evidence for a failure
of the $^3$P$_0$ model. Although this is nominally a theory summary,
hadron physics is largely driven by experiment, so I have actually cited 
some new experimental results that seemed of special interest to theorists.

I have been rather selective in this report, due primarily
to a lack of time available for completion of this summary. For
this reason many of the results presented at HADRON2001, notably
relating to heavy quark and quarkonium physics and baryon physics,
were not discussed here. The "future facilities" discussions have
clearly shown that this concentration on light $u,d,s$ hadrons will
change in future meetings, at which time we can expect to see exciting new
results on charmonium states, both regarding the states themselves and
their decay products.
The traditional concentration of the HADRON conference
series on meson physics was also discussed at this meeting, and it was 
suggested that in future there should be a serious effort to include
developments in baryons as a major part of the meeting. With new results
from facilities such as Jefferson Lab, this will certainly be appropriate,
and will make the job of the conference summary speakers even more difficult. 

\begin{theacknowledgments}
It is a great pleasure to thank Prof.Zaitsev and the organisers of HADRON2001 
for 
their kind invitation to review some of the theoretical aspects of the
physics discussed at this meeting. I would also like to thank E.Klempt 
for our collaboration in preparing our joint HADRON2001 
summary presentation, 
and
E.S.Swanson for proofreading this report.
This work was supported in part by
the DOE Division of Nuclear Physics,
at ORNL,
managed by UT-Battelle, LLC, for the US Department of Energy
under Contract No. DE-AC05-00OR22725, and by the US National Science
Foundation under Grant No. INT-0004089.
\end{theacknowledgments}


\doingARLO[\bibliographystyle{aipproc}]
          {\ifthenelse{\equal{\AIPcitestyleselect}{num}}
             {\bibliographystyle{arlonum}}
             {\bibliographystyle{arlobib}}
          }
\bibliography{summ}

\end{document}